\documentclass[10pt,onecolumn,aps,prd,floats,floatfix,superscriptaddress,nofootinbib]{revtex4}
\usepackage{amsmath,amsfonts,dsfont,amsthm,amssymb}
\usepackage[colorlinks=true, pdfstartview=FitV, linkcolor=blue, citecolor=red, urlcolor=magenta]{hyperref}
\usepackage{caption}
\usepackage{mathtools}
\usepackage{multirow}
\usepackage{graphicx}
\usepackage{color}
\usepackage{epsfig}
\usepackage{verbatim}
\usepackage{float}
\usepackage{hyperref}
\theoremstyle{plain}

\theoremstyle{definition}

\theoremstyle{remark}

\numberwithin{equation}{section}

\begin{document}

\title{\bf Feasible Regions of GB phenomenon in Extended Proca-Nuevo Gravity}
\author{Abdul Malik Sultan
\footnote{ams@uo.edu.pk; maliksultan23@gmail.com}} \affiliation{Department of Mathematics, University of Okara, Okara-56300 Pakistan.}
\affiliation{Center for Theoretical Physics, Khazar University, 41 Mehseti Street, Baku, AZ1096, Azerbaijan}
\author{Abdul Jawad \footnote{abduljawad@cuilahore.edu.pk}}
\affiliation{Department of
Mathematics, COMSATS University Islamabad, Lahore-Campus,\\
Lahore-54000, Pakistan.}
\address{Institute for Theoretical Physics
and Cosmology, Zhejiang University of Technology, Hangzhou 310023,
China.}
\author{Hamood Ur Rehman \footnote{hamood@uo.edu.pk}}
\affiliation{Department of Mathematics, University of Okara, Okara-56300 Pakistan.}
\author{N. Myrzakulov \footnote {nmyrzakulov@gmail.com}}
\affiliation{L. N. Gumilyov Eurasian National University, Astana
010008, Kazakhstan.}
\author{Sanjar Shaymatov \footnote{sanjar@astrin.uz}}\affiliation{Institute of Fundamental and Applied Research, National
Research University TIIAME, Kori Niyoziy 39, Tashkent 100000,
Uzbekistan}\affiliation{University of Tashkent for Applied Sciences,
Str. Gavhar 1, Tashkent 100149, Uzbekistan.}
\affiliation{Tashkent State Technical University, 100095 Tashkent, Uzbekistan}
\affiliation{Western Caspian University, Baku AZ1001, Azerbaijan.}

\maketitle

\section*{Abstract}
In this article, we investigate the gravitational baryogenesis mechanism in the framework of Extended Proca-Nuevo gravity, a theory where a massive vector field is non-minimally coupled to the curvature. This analysis is carried out for an early universe, encompassing three separate cosmological scenarios defined by power-law, exponential, and modified exponential scale factors. By deriving the modified field equations from the EPN action, we obtain precise solutions for each scale factor, including the influence of the vector field. We compute the baryon-to-entropy ratio using the gravitational baryogenesis formalism, where the baryon asymmetry arises from a dynamical coupling of the baryon current with the derivative of the Ricci scalar. Moreover, we discuss the stability of the assumed cosmological model in two different ways, the dynamical system analysis and squared speed of sound parameter. Our findings demonstrate that the baryon-to-entropy ratio is consistent with observational constraints in all scenarios, highlighting the potential of EPN gravity as a viable theory to explain the matter-antimatter asymmetry in the early universe. Moreover, stability analysis through dynamical system leads to a deSitter stability which is a trivial case while squared speed of sound gives that system is stable for all three scale factor models. The study further stresses the contribution of anisotropy and vector field dynamics to the cosmological evolution within modified gravity models.

\section{Introduction}

The observed disparity between matter and antimatter in the Universe is a crucial unresolved problem in cosmology and particle physics. While the Standard Model of particle physics has excelled in explaining basic interactions, it lacks a sufficient mechanism for generating the observed baryon asymmetry \cite{001,002}. This inconsistency calls for the exploration of alternative mechanisms, particularly baryogenesis scenarios, which have been the subject of extensive research \cite{003}. Gravitational baryogenesis (GB) is one of the most compelling approaches, as it links the generation of asymmetry to the gravitational sector, offering a bridge between cosmology, high-energy physics, and modified gravity \cite{004,0044}.

Baryogenesis, in its broadest sense, depends on the three essential Sakharov conditions: violation of baryon number, C (charge conjugation) and CP (charge-parity) symmetry violations, and departure from thermal equilibrium \cite{005}. Baryon number violation is indispensable, as it enables the imbalance of matter and antimatter in the universe \cite{006}. The Standard Model predicts baryon number violation via non-perturbative sphaleron processes under high-temperature conditions. However, these processes are not enough on their own; C and CP violation must also be present, since in a CP-symmetric universe, baryon and antibaryon creation would occur at identical rates, canceling out the asymmetry \cite{007}. Though CP violation exists in the Standard Model through the Cabibbo-Kobayashi-Maskawa (CKM) matrix, its magnitude is too small to account for the baryon asymmetry observed in the universe. The third Sakharov condition, prevents CPT symmetry from enforcing an equal rate of baryon and antibaryon production. This phenomenon can occur through first-order phase transitions, such as the electroweak phase transition, or other processes, like the decay of heavy particles out of equilibrium \cite{008}. 

Traditional baryogenesis frameworks, including electroweak baryogenesis and leptogenesis, introduce specific mechanisms to satisfy these conditions. Electroweak baryogenesis \cite{009,0010}, for instance, takes advantage of early universe conditions, where the electroweak phase transition could produce the required out-of-equilibrium environment. However, the transition in the Standard Model is too weak, requiring modifications like supersymmetry to enhance the phase transition and bring in extra CP-violating sources. In contrast, leptogenesis posits that the decay of heavy right-handed neutrinos in the early Universe leads to a lepton asymmetry, which is then converted into a baryon asymmetry via sphaleron interactions \cite{0011}-\cite{0014}. While these mechanisms show potential, both necessitate physics beyond the Standard Model, including new interactions, extended Higgs sectors, or particles like heavy Majorana neutrinos, to produce a baryon asymmetry in line with observations \cite{0015,0016}.

Gravitational baryogenesis distinguishes itself by considering the baryon current's interaction with the Ricci scalar or other curvature invariants, tying the matter-antimatter asymmetry to the structure of spacetime. The theory of GB posits that the time evolution of the Ricci scalar $R$ in the early universe induces CP violation, favoring the creation of baryons over antibaryons, thereby leading to the present baryon asymmetry. Gravitational baryogenesis relies solely on gravitational effects, providing a naturally embedded solution within the structure of general relativity (GR) and its extensions. The simplest implementation of GB involves a coupling of the form
\begin{equation}\nonumber
\frac{1}{M_*^2} \partial_\mu (\mathcal{I}) J^\mu,  
\end{equation}
where $M_*$ is a high-energy cutoff scale, $\mathcal{I}$ represents the gravity related invariant that drives baryogenesis, and $J^\mu$ is the baryon current. The coupling $\partial_\mu R J^\mu$ used in gravitational baryogenesis arises naturally in effective field theory such as EPN gravity. Since the EPN theory modifies the gravitational sector through a dynamical vector field, the resulting curvature scalars acquire time dependence, allowing for the lowest–dimension, CPT-violating operator that couples curvature gradients to the baryon current. This operator is fully consistent with general covariance and therefore its inclusion in baryogenesis analysis is theoretically justified. However, this approach often struggles with issues related to the magnitude of the Ricci scalar in standard cosmological evolution, leading researchers to explore alternative formulations within modified gravity paradigms. The motivation behind considering changes to GR comes from the realization that the early Universe likely involved powerful gravitational phenomena, requiring a broader description beyond Einstein’s field equations.

Extensive studies on gravitational baryogenesis have been conducted within various modified gravity frameworks, aiming to resolve the challenges poised by GR. Studies have investigated baryogenesis in $f(T)$ gravity \cite{010}, Gauss-Bonnet \cite{011}, $f(R)$ gravity \cite{012,013}, Lorentz-violating theories \cite{014}, $f(T,\Theta)$ gravity \cite{015}, $f(Q,C)$ gravity \cite{016}, $f(T,\tau)$ gravity \cite{017}, $f(R,Lm)$ gravity \cite{018}, and others \cite{019}-\cite{027}. These methods alter the gravitational sector in various ways, affecting the evolution of curvature scalars or torsion, and opening new possibilities for generating the observed baryon asymmetry. The range of these models demonstrates the persistent effort to figure out how alterations to gravity can naturally support the conditions required for baryogenesis, all while aligning with cosmological and astrophysical observations. One can also discuss this phenomenon for specific modified gravity models \cite{a}-\cite{b}.


A particularly noteworthy class of modified gravity models is represented by vector-tensor theories, where vector fields are pivotal in modifying the behavior of spacetime. Recent progress in this domain is the introduction of Proca-Nuevo gravity along with its extended formulations. Proca fields, characterized by their nonzero mass, introduce additional degrees of freedom that can influence the universe's expansion history and structure formation. In the context of the generalized Proca theory, De Felice et al. \cite{01} studied how the effective gravitational couplings for cosmological perturbations can be affected by intrinsic vector modes, potentially modifying the effective gravitational constant and leading to a phantom-like dark energy (DE) equation of state without introducing ghosts or instabilities. De Felice et al. \cite{02} further conducted an observational analysis using data from CMB, BAO, SNe Ia, and local \( H_0 \) measurements, finding a preferred value of $s = 0.254^{ -0.097}_{ +0.118}$, demonstrating that the Proca model can help alleviate the tension in the Hubble constant between high and low redshift measurements.

Expanding the scope of this framework, Nakamura  et al. \cite{03} delved into theories beyond generalized Proca, analyzing late-time cosmic acceleration and its potential observational outcomes. Their findings showed that these theories affect the cosmic growth history, where the growth rate of matter perturbations matches redshift-space distortion data, helping to distinguish DE models in beyond-generalized Proca theories from both generalized Proca theories and the $\Lambda$CDM model. A new type of Proca interaction is introduced in  \cite{04}, fulfilling a nontrivial constraint and propagating the correct degrees of freedom for a healthy massive spin-1 field, with scattering amplitudes distinct from those in Generalized Proca, thereby showing their fundamental differences.

The evolution of Proca theories has paved the way for Proca-Nuevo gravity, a nonlinear theory of a massive spin-1 field, defined by a nonlinearly realized constraint that sets it apart from traditional generalized vector models \cite{05}. Research on the quantum stability of Proca-Nuevo interactions has demonstrated that, even with differences in their classical formulations, Proca-Nuevo and Generalized Proca theories exhibit comparable behaviors at the quantum scale \cite{06}. Cosmological studies have put Proca-Nuevo gravity to the test against observational data, with analyses using SNIa and Cosmic Chronometers (CC) generating likelihood contours for the model’s free parameters \cite{07}. Further, based on analyses from DESI and BAO data, Proca-Nuevo gravity has been shown to provide robust constraints on Hubble constant \cite{08}, demonstrating a strong agreement with observational measurements. Most recently, it has been shown in \cite{09} that this gravity model is in agreement with the observed abundances of light elements formed during Big Bang Nucleosynthesis.


\begin{figure}[htbt]
\centering
\begin{minipage}[b]{0.6\textwidth}
\includegraphics[width=\textwidth]{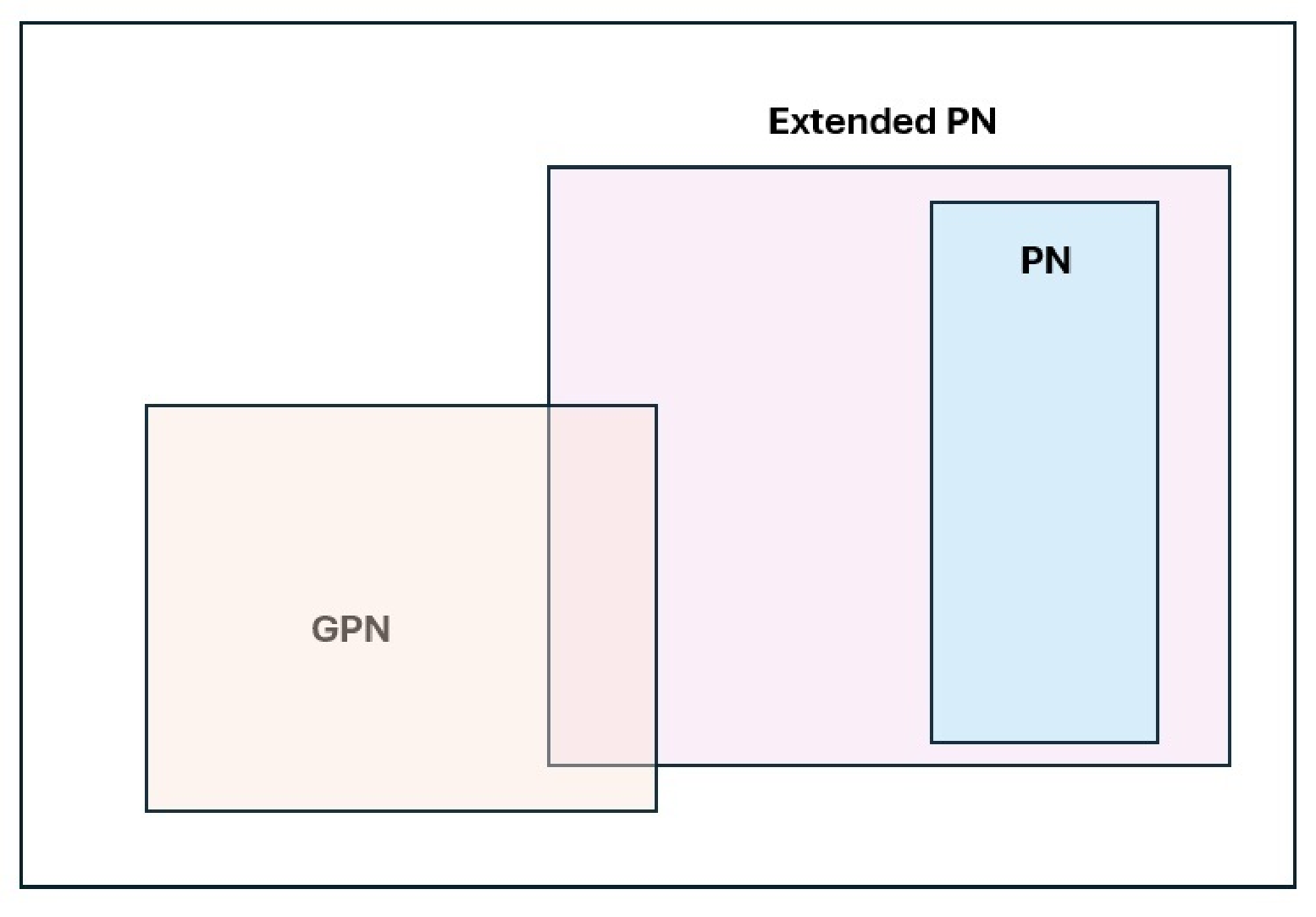}
\end{minipage}
\caption{Schematic representation of the relation between Proca-Nuevo type theories \cite{05}.}
\end{figure}

Given the rising interest in vector-tensor theories and the promising cosmological prospects of Proca-Nuevo-type models, exploring their potential to tackle core cosmological problems is both timely and significant. The consistent theoretical framework of EPN gravity and its ability to accommodate early-universe dynamics prompt us to examine its potential role in the generation of baryon asymmetry. Thus, this paper examines the implications of EPN gravity for GB. We analyze the modifications to the baryon asymmetry generation mechanism introduced by the vector field and its coupling to curvature invariants. Through an analysis of the cosmological dynamics of this theory, we aim to determine whether it provides a feasible and more robust mechanism for generating the observed baryon asymmetry of the universe. We aim to derive the relevant field equations, highlight the key CP-violating terms, and assess their effects on baryogenesis during different cosmological periods.


The arrangements of this article are as follows: in the upcoming section, an overview of EPN gravity is given which leads to the basic mathematics and field equations of the gravity. In sec III, the basic scenario of GB is discussed. In sec IV, the GB phenomenon is discussed in context of EPN gravity for different choices of scale factors. In the last section we conclude our findings.

\section{EPN Cosmology: Overview}

In this section, we will overview the EPN gravity theory which is an extension of standard Proca theory \cite{04,05}, describing a massive spin-1 field through coupling nonlinear interactions which ensure compatibility and a system free from ghost-like instabilities \cite{06}-\cite{09}. This extension is based on de Rham-Gabadadze-Tolley (dRGT) massive gravity \cite{045}, obeying its physical principles in the vector field. A profound feature of this theory which distinguishes it from other generalized vector theories is the implementation of non-linearity based constraints. Due to the covariant structure, EPN theory obtained consistent and ghost-free solutions when coupled with gravity. Such solutions represent a precise thermal history of the universe. Moreover, these solutions are suitable to predict a late-time cosmic acceleration and have a good alignment with recent observational data. The action term for this theory can be given as \cite{06}-\cite{09}
\begin{equation}\label{1}
S = \int  \sqrt{-g} \left( \frac{M_{{pl}}^2}{2} R + \mathcal{L}_{{EPN}} + \mathcal{L}_M \right)d^4x,
\end{equation}
where, $\mathcal{L}_M$ is the standard matter Lagrangian $R$ represents the Ricci scalar which is
\begin{equation}\label{2}
R=6(\dot{H}+2H^2),
\end{equation}
with $H=\frac{\dot{a}}{a}$ is Hubble parameter. The term $a=a(t)$ is scale factor of the universe and overhead dot means derivative with respect to cosmic time $t$. The massive spin-1 Lagrangian $\mathcal{L}_{{EPN}}$ can be given as
\begin{equation}\label{3}
\mathcal{L}_{{EPN}} = -\frac{1}{4} F_{\mu\nu} F^{\mu\nu} + \Lambda^4 (\mathcal{L}_0 + \mathcal{L}_1 + \mathcal{L}_2 + \mathcal{L}_3),
\end{equation}
where $\mu$ and $\nu$ represent spacetime indices taking values $0,~1,~2,~3$, labeling the temporal and spatial components of tensors in the EPN framework and 
\begin{eqnarray}\label{4}
\mathcal{L}_0 &=& a_0(X), \\\label{5}
\mathcal{L}_1 &=& a_1(X) \mathcal{L}_1[K] + d_1(X) \frac{\mathcal{L}_1[\nabla A]}{\Lambda^2}, \\\label{6}
\mathcal{L}_2 &=& \bigg[a_2(X) + d_2(X) \bigg] \frac{R}{\Lambda^2} + a_{2,X} (X) \mathcal{L}_2[K] + d_{2,X} (X) \frac{\mathcal{L}_2[\nabla A]}{\Lambda^4}, \\\label{7}
\mathcal{L}_3 &=& \left[ a_3(X) K^{\mu\nu} + d_3(X) \frac{\nabla_\mu A_\nu}{\Lambda^2} \right] \frac{G^{\mu\nu}}{\Lambda^2} - \frac{1}{6} a_{3,X} (X) \mathcal{L}_3[K] - \frac{1}{6} d_{3,X} (X) \frac{\mathcal{L}_3[\nabla A]}{\Lambda^6}.
\end{eqnarray}
Here, $A_\mu$ denotes the vector field, the term $\Lambda$ stands for the energy scale that controls the strength of vector field self-interactions. These self-interactions arise from higher-order operators in the EPN Lagrangian designed to keep the field equations second order and avoid Ostrogradsky instabilities. The coefficients $a_n(X)$ and $d_n(X)$ are functions depending on $X$ which can be given as
\begin{equation}\nonumber
  X = - \frac{1}{2\Lambda^2} A^\mu A_\mu.
\end{equation}
Here $X$ in the subscript denotes the derivative with respect to $X$. Moreover, $[K]=tr(K)$ is the trace of tensor $K^{\mu}_{\nu}=X^{\mu}_{\nu}-\delta^{\mu}_{\nu}$ with $X^{\mu}_{\nu}[A]=(\sqrt{\eta^{-1} f[A]})^{\mu}_{\nu}$ \cite{045, 046}. Here $\eta_{{\mu}{\nu}}$ represents the flat Minkowski metric and $f_{\mu \nu}[A]$ is the Stuckelberg-Lorentz tensor which is given as 
\begin{equation}\label{8}
   f_{\mu \nu}[A]= \eta_{\mu \nu}+ 2\frac{\partial_({\mu} A_{\nu)}}{\Lambda^2}+\frac{\partial_{\mu} A^{\rho} \partial_{\nu} A_{\rho}}{\Lambda^4}.
\end{equation}
Furthermore, a non-minimal coupling term proportional to $R$ is present in $\mathcal{L}_2$ and the term $G_{\mu \nu}$ appearing in $\mathcal{L}_3$ is the Einstein tensor. To apply the above theory on cosmological frame work we choose FRW metric which can be given as
\begin{equation}\label{9}
    ds^2 =- N^2(t)dt^2 + a^2(t)(dx^2 + dy^2 + dz^2),
\end{equation}
where N(t) is the lapse function which we set $N=1$. Moreover, the vector field configuration is defined as
\begin{equation}\label{10}
    A_\mu dx^\mu = -\phi(t) dt.
\end{equation}
Here, $\phi$ is a scalar field. Variation of the action in Eq. (\ref{1}) with respect to the metric leads to the Friedmann equations as
\begin{align}\label{11}
    H^2 &= \frac{1}{3M_{{Pl}}^2} (\rho_m + \rho_{{EPN}}), \\\label{12}
    \dot{H} + H^2 &= -\frac{1}{6M_{{Pl}}^2} (\rho_m + \rho_{{EPN}} + 3p_m + 3p_{{EPN}}).
\end{align}
Here, $\rho_m$ is the energy density and $p_m$ is the pressure component of the matter fluid. However, the variation with respect to scalar field $\phi(t)$ leads to the relation
\begin{align}\label{13}
    a_{0,X} +3(a_{1,X}+d_{1,X})\frac{H \phi}{\Lambda^2}=0.
\end{align}
The effective DE sector containing energy density and pressure in the frame work of EPN cosmology can be given as
\begin{align}\label{14}
    \rho_{{EPN}} &= \Lambda^4 \left( -a_0 + a_{0,X} \frac{\phi^2}{\Lambda^2} + 3(a_{1,X} + d_{1,X}) \frac{H \phi^3}{\Lambda^4} \right), \\\label{15}
    p_{{EPN}} &= \Lambda^4 \left( a_0 - (a_{1,X} + d_{1,X}) \frac{\phi^2 \dot{\phi}}{\Lambda^4} \right).
\end{align}
An interesting aspect of this scenario is that the vector field equation, and consequently due to the ansatz (\ref{10}), Eq. (\ref{13}) having scalar field represents a non-dynamical system which contributes as constraint that develop an algebraic relation between scalar field and Hubble constant. Due to the specific formulation of action (\ref{1}), the Friedmann equation completely depend on Hubble parameter. Thus the effective DE density and pressure components are given as
\begin{align}\label{16}
\rho_{{EPN}} &= \frac{1}{2} \Lambda^4 c_m y^{2/3} \left( \frac{\Lambda^4}{M_{{Pl}}^2 H^2} \right)^{1/3}, \\\label{17}
p_{{EPN}} &= 3 M_{{Pl}}^2 H^2 \left( -1 - \frac{2\dot{H}}{3H^2} \right).
\end{align}
Here \( y = 4\sqrt{\frac{6}{c_m}} \) with \( c_m \equiv \frac{m^2 M_{{Pl}}^2}{\Lambda^4} \sim 1 \) \cite{05}.
Finally, these equations close with assuming matter conservation equation as
\begin{align}\label{18}
    \dot{\rho}_m + 3H(\rho_m + p_m) &= 0,\\\label{18a}\dot{\rho}_r + 3H(\rho_r + p_r) &= 0.
\end{align}
Taking into account the Friedmann equations (\ref{11}) and (\ref{12}), the above equation leads to the DE conservation equation in EPN gravity as
\begin{align}\label{19}
       \dot{\rho}_{{EPN}} + 3H(\rho_{EPN} + p_{EPN}) =0.
\end{align}

\section{Gravitational Baryogenesis: General scenario}

The discussion points out that both the BBN predictions \cite{1} and CMB observational data \cite{2} validate the fact that the universe is dominated by matter rather than antimatter. In addition, no interactions take place that result in radiation from the annihilation of matter and antimatter. This further bolsters the hypothesis of a matter-antimatter imbalance. As per BBN measurements \cite{1}, the ratio is specified as
$\frac{\eta_{B}}{s}=(5.6 \pm 0.6)\times 10^{-10}$, while CMB \cite{2}
measures it as $\frac{\eta_{B}}{s}=(6.19 \pm 0.14)\times
10^{-10}$. This asymmetry in more detailed form is explained in  \cite{45, 46}. Drawing from the most up-to-date observational data, the indicated measure of this asymmetry also called baryon to entropy ratio (BER) is \cite{41, 42}
\begin{eqnarray}\label{20}
\frac{\eta_{B}}{s}\simeq 9.42 \times 10^{-11}.
\end{eqnarray}
As discussed in \cite{43}, the gravitational baryogenesis mechanism is based on a well-established Sakharov criterion for baryogenesis, specifically the existence of a CP-violating interaction, which usually takes the form
\begin{eqnarray}\label{21}
\frac{1} {M^{2}_{*}} \int d^{4}x\big(\partial_{\mu} \mathcal{I} \big)J^{\mu} \sqrt{-g},
\end{eqnarray}
Here, $g$ refers to the metric determinant, $R$ to the Ricci
scalar,and $M_{*}$ stands for the cutoff scale of the fundamental
effective theory \cite{44}. $J^{\mu}$ is the symbol for the baryonic current. A critical parameter, the baryon asymmetry factor, symbolized by $\eta_{B}$, plays a central role in determining baryogenesis.
\begin{eqnarray}\label{22} 
\eta_{B}&=&\eta_{\beta}-{\eta}_{\bar\beta},
\end{eqnarray}
For the interaction term to generate baryogenesis, thermal equilibrium is considered essential. When the temperature drops below the critical threshold $T_D$ the contributions to baryon asymmetry are locked in place within the framework of GB.
The mathematical form of the resulting BER is
\begin{eqnarray}\label{23}
\frac{\eta_{B}}{s} \simeq -\frac{15g_{b}}{4\pi^2g_{*}}\bigg(\frac{\dot{\mathcal{I}}}{M_{*}^{2}\tau}\bigg)\bigg|T_{D},
\end{eqnarray}
where $g_{*}$ represents the total freedom degrees associated with massless particles, while $g_b$ represents those related to baryons. The energy density $\rho$ and temperature $T$ are related in thermal equilibrium as follows:
\begin{eqnarray}\label{24}
\rho(T)&=&\frac{\pi^{2}}{30}g_{*}T_{D}^{4}.
\end{eqnarray}

\section{Gravitational Baryogenesis in EPN Gravity}

This section provides a discussion of gravitational baryogenesis in the framework of EPN cosmology. In this assumed setup, the CP-violating interaction term is proportional to Ricci scalar $R$ which can be expressed as
\begin{eqnarray}\label{25}
\frac{1}{M_*}\int\sqrt{-g}d^4x(\partial_\mu (R))j^\mu.
\end{eqnarray}
For such kind of baryogenesis interaction, the term $\frac{\eta_{B}}{s}$  can be expressed as
\begin{eqnarray}\label{26}
\frac{\eta_{B}}{s} \simeq -\frac{15g_{b}}{4\pi^2g_{*}}\bigg(\frac{\dot{R}}{M_{*}^{2}\tau}\bigg)|T_{D},
\end{eqnarray}
Considering the framework of EPN gravity, we have considered various scale factor models to analyze the gravitational baryogenesis scenario which are

\subsection{Power Law Scale Factor Model}

The power law model is the simplest scale factor with analytical manageability. It is easy to obtain analytical solution of field equations and closed-form expressions for the BER through this choice of scale factor. Moreover, such expansions effectively
describe important cosmological eras such as the radiation-dominated
era, and the matter-dominated era. These epochs are particularly
relevant for baryogenesis, which is believed to occur at early
times in the universe. In power law scale factor model, the scale factor $a(t)$ evolves as a power of cosmic time $t$ which can be given as \cite{025,055}
\begin{eqnarray}\label{27}
a(t)=a_{o} t^{p},
\end{eqnarray}
where $a_0$ is a real constant representing scale factor at some reference point, $p$ denotes the power law exponent which is used to calculate the expansion rate of the universe. For the above given scale factor, the Hubble parameter can be given as
\begin{eqnarray}\label{28}
H=\frac{p}{t}.
\end{eqnarray}
Simplifying the relation for energy density given in Eq. (\ref{16}) by inserting the values from Eqs. (\ref{2}), (\ref{27}) and (\ref{28}), we get
\begin{eqnarray}\label{29}
\rho_{EPN}&=&2^{2/3} ~\sqrt[3]{3}~ \Lambda ^4~ \sqrt[3]{\frac{\Lambda ^4 t^2}{M_{pl}^2 n^2}}~c_m^{2/3}.
\end{eqnarray}
Comparison of the above equation with Eq. (\ref{24}) can express decoupling time $t_D$ as a function of decoupling temperature $T_D $ which can be given as
\begin{eqnarray}\label{30}
t_D&=&\frac{\pi ^3~ g_*^{3/2}~ M_{pl}~ n~ T_D^{3/2}}{180 \sqrt{10}~ c_m ~\Lambda ^8}.
\end{eqnarray}
The net BER for this model can be determined by simplifying Eq. (\ref{23}) by using Eqs. (\ref{29}) and (\ref{30}) which leads to the relation
\begin{eqnarray}\label{31}
\frac{\eta_B}{s}&=&\frac{2624400000 ~\sqrt{10}~ c_m^3~ g_b ~\Lambda ^{24} (2 n-1)}{\pi^{11}~ g^{11/2}~ M_*^2 ~M_{pl}^3~ n^2~ T_D^{11/2}}.
\end{eqnarray}
\begin{figure}
\centering
\begin{minipage}[b]{0.85\textwidth}
\includegraphics[width=\textwidth]{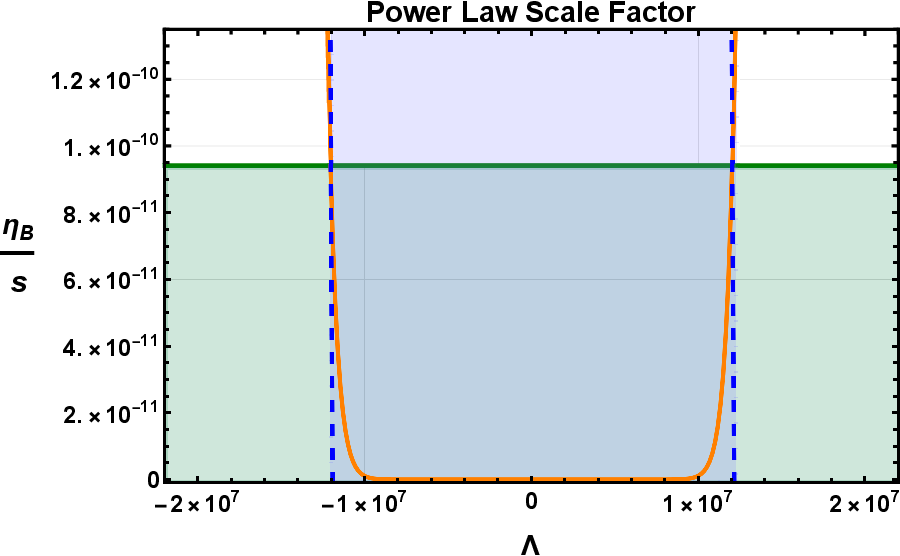}
\end{minipage}
\caption{Variation of baryon to entropy $\frac{\eta_B}{s}$ against parameter $\Lambda$, for the power law scale factor.}
\end{figure}
Using Eq. (\ref{31}), the ratio $\frac{\eta_B}{s}$ has plotted in \textbf{FIG. 2} against the parameter $\Lambda$. The analysis adopts the following parameters and physical constants: baryons have $g_b \sim \mathcal{O}(1) = 1$ internal degrees of freedom, and the decoupling temperature is set to $T_D = M_I = 2 \times 10^{16}$ GeV, with $M_I$ representing the upper bound on the amplitude of tensor mode fluctuations at the inflationary scale, inferred from LIGO's gravitational wave data. Moreover, the effective number of relativistic degrees of freedom for massless particles is taken as $g_* = 106$, consistent with the Standard Model at high energies. The Planck mass is specified as $M_P = 1.22 \times 10^{19}$ GeV, and the reduced Planck mass is given by $M_{Pl} = \frac{M_P}{\sqrt{8 \pi}}$. These values are utilized to numerically assess the evolution of $\frac{\eta_B}{s}$ as a function of $\Lambda$, enabling a direct comparison with observational limits and helping to validate the EPN gravity model in the context of baryogenesis. In this analysis, the model parameter $n$ is fixed at 0.6 to explore a specific regime of the theory. The green solid line on the plot denotes the observed BER, which is approximately $9.42\times 10^{-11}$, as determined by current cosmological observations. This serves as a point of comparison for evaluating the model's plausibility. The blue dashed line outlines the validation, representing the range where the theoretical predictions for the BER are in agreement with observational data. The plot notably shows that the EPN gravity model produces values of $\frac{\eta_B}{s}$ within the observationally valid range for $\Lambda$ values between approximately $-1.2 \times 10^7$ and $1.2 \times 10^7$. This agreement between theory and observation confirms that the EPN gravity model with a power-law scale factor and $n=0.6$ effectively explains the matter–antimatter asymmetry. Moreover, for the power-law scale factor model, the only free parameter is the power index $n$, and hence BER depends only on \(n\). Therefore, a focused sensitivity analysis was performed with respect to this parameter. We varied \(n\) about the fiducial value \(n=0.6\) within the interval \(n\in(0.5,1)\), which covers the relevant cosmological behaviors for the chosen ansatz. Throughout the range tested, the qualitative shape of the curve shown in FIG.~2 is preserved and the magnitude of \(\eta_B/s\) remains of the same order as the observed baryon asymmetry. Variations in \(n\) primarily shift the amplitude and width of the allowed region in \(\Lambda\). It is observed that \(\eta_B/s\) changes abruptly when $n\leq0.5$. Meanwhile, at $n=0.5$ baryon to entropy ratio becomes zero while at $n<0.5$ gives negative values. These checks confirm that the baryogenesis results for the power-law model are robust against moderate parameter variations when $n>0.5$ and are not driven by numerical artifacts. Furthermore, the sensitivity analysis of parameter $n$ for the power law scale factor model in EPN gravity is given in TABLE. \ref{tab:1} as
\begin{table*}[htbp]
\centering
\caption{Sensitivity analysis for power law scale factor model in EPN gravity for free parameters.}
\label{tab:1}
\begin{tabular}{p{1.1cm}| p{5cm} p{4.2cm} p{5.3cm}}
\hline
\hline
~~~\textbf{$n$} & \textbf{~~~~~~~~~~~~~$\Lambda$} & ~~~~~~~~~~~~\textbf{\(\frac{\eta_B}{s}\)} &~~~~~~~~~~~~ \textbf{Remarks} \\
\hline
\hline
~~1.0 & ~ $(-1.1775 \times 10^7,~1.1775 \times 10^7)$ & $0<\frac{\eta_B}{s}<9.42\times10^{-11}$ & Consistent with the observations \\
\hline
~~0.9 & ~ $(-1.1815 \times 10^7,~1.1815 \times 10^7)$ & $0<\frac{\eta_B}{s}<9.42\times10^{-11}$ & Consistent with the observations \\
\hline
~~0.8 & ~ $(-1.1835 \times 10^7,~1.1835 \times 10^7)$ & $0<\frac{\eta_B}{s}<9.42\times10^{-11}$ & Consistent with the observations \\
\hline
~~0.7 & ~ $(-1.1885 \times 10^7,~1.1885 \times 10^7)$ & $0<\frac{\eta_B}{s}<9.42\times10^{-11}$ & Consistent with the observations \\
\hline
~~0.6 & ~ $(-1.2075 \times 10^7,~1.2075 \times 10^7)$ & $0<\frac{\eta_B}{s}<9.42\times10^{-11}$ & Consistent with the observations \\
\hline
~~0.5 & ~~~~~~~~ For any value & ~~~~~~~Equal to zero & Inconsistent with the observations \\
\hline
~~$<0.5$ & ~~~~~~~~ For any value & ~~~~~~~Less then zero & Inconsistent with the observations \\
\hline
\hline
\end{tabular}
\end{table*}
The TABLE. \ref{tab:1} shows that the baryogenesis results for the power law scale factor model are stable against various values of the parameter in the obtained interval.


\subsection{Exponential Scale Factor Model}

This scale factor represents a cosmological scenario having a smooth transition from a quasi-static early universe to a late-time exponential expansion. The mathematical formalism of this scale factor can be given as \cite{056}
\begin{eqnarray}\label{32}
a(t)=a_0 \left(1+e^{\lambda  t}\right),
\end{eqnarray}
where $a_0$ is a real parameter used to set the initial scale while $\lambda >0$ is a constant controlling the expansion rate of the universe. Moreover, if \( t \to -\infty \), the exponential term vanishes which leads to a constant scale factor $a(t) \approx a_0$ which means a non-singular universe originating from a static or quasi-static state. For the late time behavior when \( t \to +\infty \implies e^{\lambda t} \gg 1  \), thus the scale factor evolves as $ a(t) \approx a_0 e^{\lambda t}$ which represents the exponential expansion mimicking a de Sitter universe and thus consistent with accelerated cosmic expansion driven by DE. The Hubble parameter for this scale factor can be given as
\begin{eqnarray}\label{33}
H=\frac{\lambda  e^{\lambda  t}}{e^{\lambda  t}+1}.
\end{eqnarray}
Simplification of the relation for energy density given in Eq. (\ref{16}) by inserting the values from Eqs. (\ref{2}), (\ref{32}) and (\ref{33}) yields
\begin{eqnarray}\label{34}
\rho_{EPN}&=&2^{2/3} \sqrt[3]{3} \Lambda ^4  c_m^{2/3} \sqrt[3]{\frac{\Lambda ^4 e^{-2 \lambda  t} \left(e^{\lambda  t}+1\right)^2}{\lambda ^2 M_{pl}^2}} .
\end{eqnarray}
Comparison of the above equation with Eq. (\ref{24}) can express decoupling time $t_D$ as a function of decoupling temperature $T_D $ which can be given as
\begin{eqnarray}\label{35}
t_D&=&\lambda^{-1} \log \left(\frac{180 \left(\pi ^3~ \sqrt{10}~ c_m~ g_*^{3/2} \lambda ~ \Lambda ^8~ M_{pl}~ T_D^{3/2}+1800 c_m^2~  \Lambda^{16}\right)}{\pi ^6~ g_*^3~ \lambda ^2~ M_{pl}^2~ T_D^3-324000 c_m^2 ~\Lambda ^{16}}\right).
\end{eqnarray}
The net BER for the exponential scale factor model can be obtained by simplifying Eq. (\ref{23}) by using Eqs. (\ref{34}) and (\ref{35}) which leads to the relation
\begin{eqnarray}\nonumber
\frac{\eta_B}{s}&=& -\big(\pi^{11} ~g_*^{11/2}~ M_*^2 ~M_{pl}^3 T_D^{11/2}\big)^{-1} \bigg[4050 c_m ~ g_b~ \Lambda ^8 (-648000 \sqrt{10} ~c_m^2~ \Lambda ^{16} \\ \label{36} &+& 1800 \pi ^3 ~c_m~ g_*^{3/2}~ \lambda~  \Lambda ^8~ M_{pl}~ T_D^{3/2}+\pi ^6~ \sqrt{10}~ g_*^3 ~\lambda ^2 M_{pl}^2 T_D^3)\bigg].
\end{eqnarray}
\begin{figure}
\centering
\begin{minipage}[b]{0.85\textwidth}
\includegraphics[width=\textwidth]{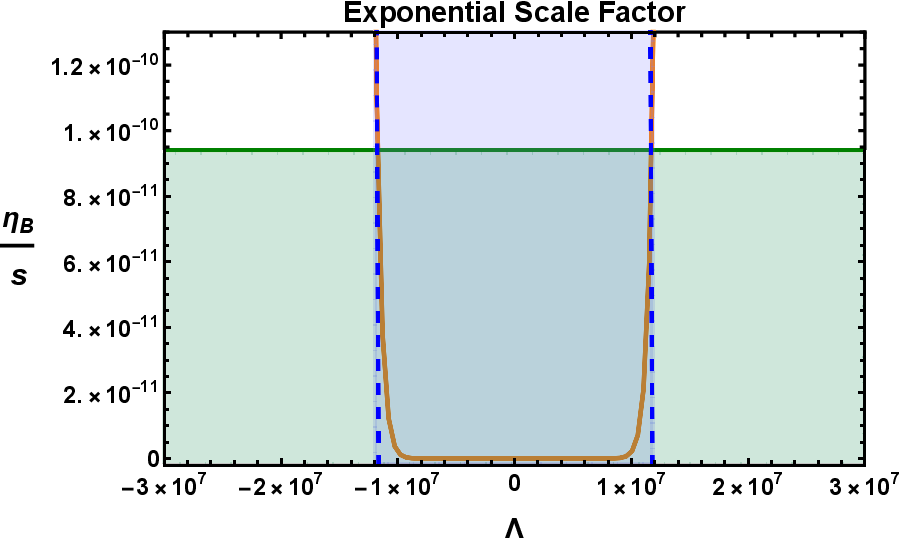}
\end{minipage}
\caption{Variation of BER $\frac{\eta_B}{s}$ against parameter $\Lambda$, in EPN cosmology for exponential scale factor.}
\end{figure}
In \textbf{FIG. 3}, the variation of the baryon-to-entropy ratio $\frac{\eta_B}{s}$ with respect to $\Lambda$ is depicted using Eq. (\ref{36}), with $\lambda$ set to 0.3. This figure adopts the same analysis framework as FIG. 2, where the solid green line denotes the observational value of $\frac{\eta_B}{s} \approx 9.42 \times 10^{-11}$. The dashed blue lines highlight the interval of parameter $\Lambda$ in which BER lies in the observational bounds. The plot indicates that the theoretical predictions for $\eta_B/s$ agree with the observational data within the range $\Lambda \in (-1.17 \times 10^7, 1.17 \times 10^7)$. The model successfully reproduces the observed matter–antimatter asymmetry within this validity region, enhancing the viability of EPN gravity as a mechanism for baryogenesis under the specified conditions. As in the previous case, the same physical assumptions are used: $g_b = 1$, $T_D = 2 \times 10^{16}$ GeV, $g_* = 106$, and Planck mass values $M_P = 1.22 \times 10^{19}$ GeV and $M_{Pl} = \frac{M_P}{\sqrt{8 \pi}}$. The results remain consistent, emphasizing the strong predictive capacity of the model in various dynamical situations. Moreover, in case of exponential scale factor the BER ratio \(\eta_B/s\) involves only the parameter \(\lambda\). To examine the robustness of our results and sensitivity of the parameters, we varied \(\lambda\) around the fiducial value \(\lambda=0.3\) within the range \(\lambda\in[0.1,0.6]\). The resulting \(\eta_B/s\) curves retain the same qualitative structure as shown in FIG. 3 with no changes in the amplitude and width of the viable \(\Lambda\)-interval. Furthermore, the sensitivity analysis of parameter $\lambda$ for the exponential scale factor model in EPN gravity is given in TABLE. \ref{tab:2} as
\begin{table*}[htbp]
\centering
\caption{Sensitivity analysis for exponential scale factor model in EPN gravity for free parameters.}
\label{tab:2}
\begin{tabular}{p{1.1cm}| p{5cm} p{4.2cm} p{5.3cm}}
\hline
\hline
~~~\textbf{$\lambda$} & \textbf{~~~~~~~~~~~~~$\Lambda$} & ~~~~~~~~~~~~\textbf{\(\frac{\eta_B}{s}\)} &~~~~~~~~~~~~ \textbf{Remarks} \\
\hline
\hline
~~0.6 & ~ $(-1.1710 \times 10^7,~1.1710 \times 10^7)$ & $0<\frac{\eta_B}{s}<9.42\times10^{-11}$ & Consistent with the observations \\
\hline
~~0.5 & ~ $(-1.1710 \times 10^7,~1.1710 \times 10^7)$ & $0<\frac{\eta_B}{s}<9.42\times10^{-11}$ & Consistent with the observations \\
\hline
~~0.4 & ~ $(-1.1710 \times 10^7,~1.1710 \times 10^7)$ & $0<\frac{\eta_B}{s}<9.42\times10^{-11}$ & Consistent with the observations \\
\hline
~~0.3 & ~ $(-1.1795 \times 10^7,~1.1795 \times 10^7)$ & $0<\frac{\eta_B}{s}<9.42\times10^{-11}$ & Consistent with the observations \\
\hline
~~0.2 & ~ $(-1.1795 \times 10^7,~1.1795 \times 10^7)$ & $0<\frac{\eta_B}{s}<9.42\times10^{-11}$ & Consistent with the observations \\
\hline
~~0.1 &  ~ $(-1.1795 \times 10^7,~1.1795 \times 10^7)$  & $0<\frac{\eta_B}{s}<9.42\times10^{-11}$  & Consistent with the observations \\
\hline
\hline
\end{tabular}
\end{table*}
The TABLE. \ref{tab:2} also demonstrates that the baryogenesis results for the exponential scale factor are stable against various values of the parameter in the obtained interval.


\subsection{Modified Exponential Scale Factor Model}

The modified exponential scale factor provides a dynamically rich and analytically manageable background where curvature quantities evolve non-trivially with cosmic time $t$. It is typically developed to ensure the cosmic acceleration whose mathematical formalism can be given as
\begin{eqnarray}\label{37}
a(t)=a_{o} e^{\alpha t+\beta t^2},
\end{eqnarray}
where $a_0$ is a real constant representing scale factor at some reference point, $\alpha,~\beta$ also represent real constants with $\beta>0$ typically assumed to ensure accelerated expansion. Moreover, at early times  (\( t \to 0 \)), the non-linear term $\beta t^2$ is subdominant, and thus scale factor behaves like $a(t)=a_0~e^{\alpha t}$ resembling with standard exponential scale factor. At late times, the term $\beta t^2$ become dominated and hence scale factor grows much faster than simple exponential expansion. The Hubble parameter for the above mentioned scale factor can be given as
\begin{eqnarray}\label{38}
H=\alpha +2 \beta  t.
\end{eqnarray}
Simplifying the relation for energy density given in Eq. (\ref{16}) by inserting the values from Eqs. (\ref{2}), (\ref{37}) and (\ref{38}). We get
\begin{eqnarray}\label{39}
\rho_{EPN}&=&2^{2/3}~ \sqrt[3]{3}~ \Lambda ^4~c_m^{2/3} \sqrt[3]{\frac{\Lambda ^4}{M_{pl}^2 (\alpha +2 \beta  t)^2}}.
\end{eqnarray}
Comparison of the above equation with Eq. (\ref{24}) can express decoupling time $t_D$ as a function of decoupling temperature $T_D $ which can be given as
\begin{eqnarray}\label{40}
t_D&=&\frac{\pi ^3 \alpha  \left(-g_*^{3/2}\right) M_{pl}~ T_D^{3/2}-180 \sqrt{10}~ c_m~ \Lambda ^8}{2 \pi ^3 \beta  g_*^{3/2} M_{pl} T^{3/2}}.
\end{eqnarray}
The net BER for this model can be determined by simplifying Eq. (\ref{23}) by using Eqs. (\ref{39}) and (\ref{40}), which leads to the relation
\begin{eqnarray}\label{41}
\frac{\eta_B}{s}&=&\frac{32400 \sqrt{10} ~\beta ~ c_m ~g_b ~\Lambda ^8}{\pi^5~ g_*^{5/2}~ M_*^2~ M_{pl}~ T_D^{5/2}}.
\end{eqnarray}
\begin{figure}
\centering
\begin{minipage}[b]{0.85\textwidth}
\includegraphics[width=\textwidth]{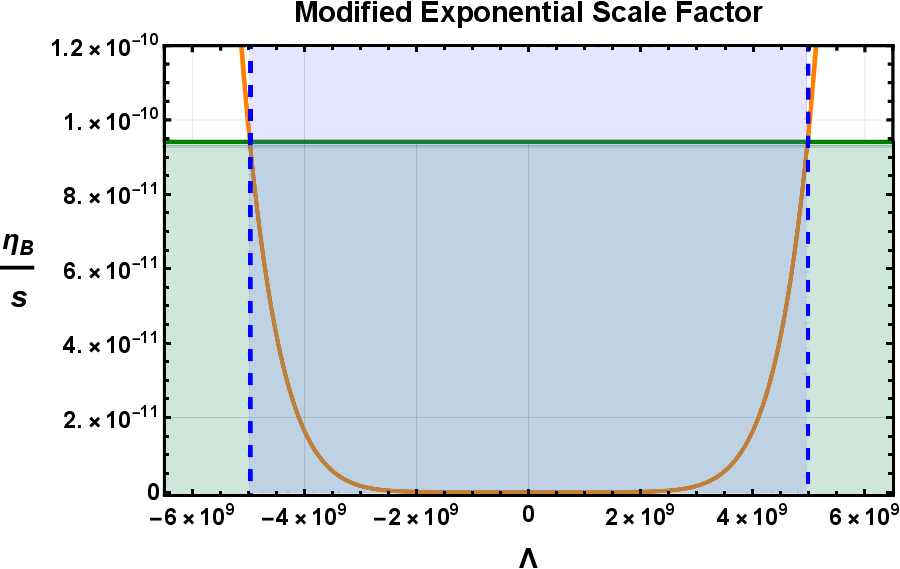}
\end{minipage}
\caption{Variation of $\frac{\eta_B}{s}$ against parameter 
$\Lambda$, for modified exponential scale factor.}
\end{figure}
The ratio $\frac{\eta_B}{s}$, derived from Eq. (\ref{41}), is displayed in \textbf{FIG. 4} as a function of the parameter $\Lambda$. Here, $\beta$ is held constant at 0.3. The plot structure remains unchanged, with the solid green line representing the observational baryon asymmetry value and the dashed blue lines outlining the acceptable observational range. The validity region in this scenario, where the model’s predictions remain within observational limits, is much wider, spanning $\Lambda \in (-5 \times 10^9, 5 \times 10^9)$. The extension of the $\Lambda$ range indicates that the model shows enhanced flexibility or stability in replicating the observed baryon asymmetry. Moreover, for the modified exponential scale factor,
the BER \(\eta_B/s\) explicitly depends on the model parameter \(\beta\). In the present analysis, we chose \(\beta=0.3\) as a representative value. To study the sensitivity of our results, we varied \(\beta\) in the range \(\beta\in[0.1,0.7]\). The resulting behavior of \(\eta_B/s\) remains qualitatively unchanged, with only a modest reduction in the width of the viable \(\Lambda\)-interval for increasing \(\beta\). The amplitude of \(\eta_B/s\) consistently lies around \(10^{-10}\), consistent with the observed baryon asymmetry. Furthermore, the sensitivity analysis of parameter $\beta$ for the modified exponential law scale factor model in EPN gravity is given in TABLE. \ref{tab:3} as
\begin{table*}[htbp]
\centering
\caption{Sensitivity of free parameters for modified exponential scale factor model in EPN gravity.}
\label{tab:3}
\begin{tabular}{p{1.1cm}| p{5cm} p{4.2cm} p{5.3cm}}
\hline
\hline
~~~\textbf{$\beta$} & \textbf{~~~~~~~~~~~~~$\Lambda$} & ~~~~~~~~~~~~\textbf{\(\eta_B/s\)} &~~~~~~~~~~~~ \textbf{Remarks} \\
\hline
\hline
~~0.7 & ~ $(-4.4920 \times 10^9,~4.4920 \times 10^9)$ & $0<\frac{\eta_B}{s}<9.42\times10^{-11}$ & Consistent with the observations \\
\hline
~~0.6 & ~ $(-4.5634 \times 10^9,~4.5634 \times 10^9)$ & $0<\frac{\eta_B}{s}<9.42\times10^{-11}$ & Consistent with the observations \\
\hline
~~0.5 & ~ $(-4.6883 \times 10^9,~4.6883 \times 10^9)$ & $0<\frac{\eta_B}{s}<9.42\times10^{-11}$ & Consistent with the observations \\
\hline
~~0.4 & ~ $(-4.8132 \times 10^9,~4.8132 \times 10^9)$ & $0<\frac{\eta_B}{s}<9.42\times10^{-11}$ & Consistent with the observations \\
\hline
~~0.3 & ~ $(-4.9917 \times 10^9,~4.9917 \times 10^9)$ & $0<\frac{\eta_B}{s}<9.42\times10^{-11}$ & Consistent with the observations \\
\hline
~~0.2& ~ $(-5.2415 \times 10^9,~5.2415 \times 10^9)$ & $0<\frac{\eta_B}{s}<9.42\times10^{-11}$ & Consistent with the observations \\
\hline
~~$0.1$ & ~ $(-5.7233 \times 10^9,~5.7233 \times 10^9)$ & $0<\frac{\eta_B}{s}<9.42\times10^{-11}$ & Consistent with the observations \\
\hline
\hline
\end{tabular}
\end{table*}
The TABLE. \ref{tab:3} shows that the baryogenesis results for the modified exponential scale factor model are both numerically stable and physically robust against reasonable parameter variations in context of EPN gravity.

\section{Stability Analysis}

It is mandatory for any vector–tensor-modified gravity theory to be free from ghost and tachyon instabilities to remain physically viable. In this regard, dynamical system analysis is a useful tool for investigating the qualitative behavior of EPN cosmological model and identifying the critical points of the underlying equations of motion. These critical points can be obtained from some dimensionless variables that lead to a system of differential equations. The eigenvalues are calculated to discuss the stability condition for each critical point of the linearized system. Here we consider radiation dominated era case for which continuity equations become as \cite{05}
\begin{eqnarray}\nonumber
p_m=0 \quad \Rightarrow  \quad \dot{\rho}_m + 3H\rho_m = 0,\quad
p_r=\frac{1}{3} \rho_r \quad \Rightarrow  \quad \dot{\rho}_r + 4H\rho_r = 0
\end{eqnarray}
To examine the dynamical stability of EPN gravity, we transform Eqs. (\ref{14}) and (\ref{15}) into a system of differential equations employing the following dimensionless variables
\begin{eqnarray}\label{3.1}
x=\Omega_m=\frac{\rho_m}{3M_{pl}^2H^2},\quad y=\Omega_{r}=\frac{\rho_{r}}{3M_{pl}^2H^2},
\quad \Omega_{EPN}=\frac{\rho_{EPN}}{3M_{pl}^2H^2},
\end{eqnarray}
where the terms $x$ and $y$ must be positive in the range $0\leq x,y
\leq 1$, $\Omega_{EPN}$ is EPN component of the dark
sector. Inserting these variables in Eq. (\ref{11}) leads to the
following mathematical relation \cite{05}
\begin{eqnarray}\label{3.2}
\Omega_m+\Omega_{r}=1-\Omega_{EPN}.
\end{eqnarray}
The simplification of Eqs. (\ref{11}), (\ref{18}), (\ref{19}) and
(\ref{3.2}) leads to the following mathematical relation
\begin{eqnarray}\label{3.3}
\frac{\dot{H}}{H^2}=-\frac{3}{3+\Omega_{EPN}}\bigg(\frac{3}{2} \Omega_m+2\Omega_r\bigg).
\end{eqnarray}
Differentiating $x$ and $y$ in Eq. (\ref{3.1}) with respect to the
$e$-folds number $N=\ln a$ of the universe and
using the value of $\frac{\dot{H}}{H^2}$ from Eq. (\ref{3.3})
, one can get
equations as
\begin{eqnarray}\label{3.6}
x'&=& \frac{3x(4x+5y-4)}{4-x-y} ,\\\label{3.7}
y'&=& \frac{y(13x+16y-16)}{4-x-y} .
\end{eqnarray}
We will use this system of first order differential equations given in Eqs. (\ref{3.6}) and (\ref{3.7}) to find the critical points and corresponding eigen values for stability analysis.
\begin{figure}
\centering
\begin{minipage}[b]{0.45\textwidth}
\includegraphics[width=\textwidth]{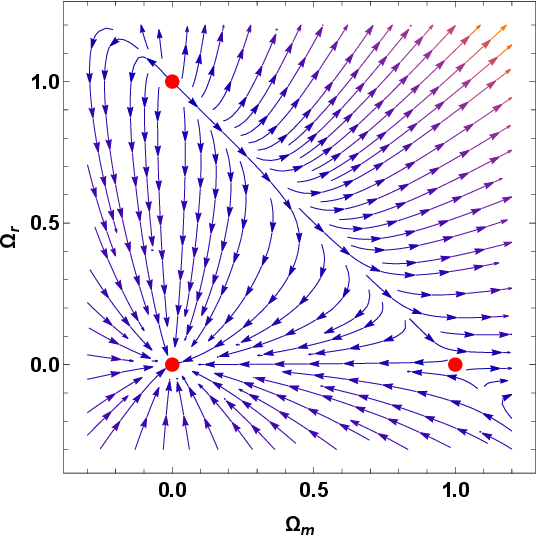}
\end{minipage}
\caption{Phase portrait associated to the autonomous system given in Eqs. (\ref{3.6}) and (\ref{3.7}). The radiation, matter and dark energy fixed points are indicated. We acknowledge the  work done by C. de Rham et al \cite{05} to discuss the dynamical stability of EPN gravity with similar outcomes.}
\end{figure}
\begin{table*}[htbp]
\centering
\caption{Critical points of the autonomous system given in Eqs. (\ref{3.6}) and (\ref{3.7}) representing the density parameters in radiation, matter and Proca fields respectively.}
\label{tab:comparison-lit}
\begin{tabular}{p{2.4cm} p{1.4cm} p{1.4cm} p{1.4cm} p{2.8cm} p{3cm}}
\hline
\hline
\textbf{~~~~~Era} & \textbf{~~~~$\Omega_r$} & ~~~\textbf{$\Omega_m$} &~~ \textbf{$\Omega_{epn}$}&~~~\textbf{Eigen Values}&~~~~\textbf{Remarks} \\
\hline
\hline
~~~Radiation &~~~~ $1$ &~~~~$0$ & ~~~~~$0$ &~~~~~~~~$4,-1$&~~~Saddle Point\\
\hline
~~~Matter &~~~~ $0$ & ~~~~$1$ &~~~~ $0$  &~~~~~~~~$\frac{16}{3},~1$&~~~~~~Unstable\\
\hline
~~~de Sitter & ~~~~~$0$ & ~~~~$0$ & ~~~~~$1$ &~~~~~~$-4,~-3$&~~~~~~~~Stable \\
\hline
\hline
\end{tabular}
\end{table*}\\
\textbf{FIG. 5} represents the phase portrait of the autonomous system given in Eqs. (\ref{3.6}) and (\ref{3.7}) in the $(\Omega_m,\Omega_r)$ plane. Moreover, the corresponding critical points and their stability has summarized in \textbf{TABLE IV}. The radiation-dominated point appears as a saddle point, containing mixed eigenvalues, matter-dominated point is unstable, indicating a transient evolutionary phase and in contrast, the trivial solution $(0,0,1)$, corresponding to the de Sitter era, possesses negative eigenvalues and acts as a stable late-time attractor \cite{05}. The phase trajectories in FIG. 5 confirm this behavior, showing that the system generically evolves from radiation and matter dominance toward the stable de Sitter state which are just a trivial case of stability. We acknowledge the  work done by C. de Rham et al \cite{05} to discuss the dynamical stability of EPN gravity with similar outcomes.

Furthermore, we consider an other way to test the stability of the EPN cosmological system by analyzing linear perturbations around a given background spacetime called squared speed of sound, \( c_s^2 \), which governs the propagation of scalar perturbations. For a model to be physically viable, the condition \( c_s^2 \ge 0 \) must be satisfied to avoid instabilities, while unusually large values of \( c_s^2 \) may indicate super-luminal modes. Therefore, the behavior of the squared speed of sound serves as a stability criterion and provides insight into the consistency of a cosmological system. It is defined as \cite{v}
\begin{eqnarray}\label{18cs}
C_{s}^{2}=\frac{\partial p}{\partial \rho}=\frac{\partial
p}{\partial t} \frac{\partial t}{\partial \rho}.
\end{eqnarray}

\begin{itemize}
\item For the power law scale factor model given in Eq. (\ref{27}), the field Eqs. (\ref{16}) and (\ref{17}) becomes as
    \begin{align}\label{5.9}
\rho_{{EPN}} &= \frac{1}{2} c_m \Lambda ^4 y^{2/3} \sqrt[3]{\frac{\Lambda ^4 t^2}{M_{pl}^2 p^2}}, \\\label{5.10}
p_{{EPN}} &= \frac{M_{pl}^2 (2-3 p) p}{t^2}.
\end{align}
From the above Eqs. (\ref{5.9}) and (\ref{5.10}), the squared speed of sound parameter can be given as
\begin{eqnarray}\label{5.11}
C_{s}^{2}=\frac{6 (3 p-2)}{c_m p y^{2/3} \left(\frac{\Lambda ^4 t^2}{M_{pl}^2 p^2}\right)^{4/3}}.
\end{eqnarray}
\begin{figure}
\centering
\begin{minipage}[b]{0.5\textwidth}
\includegraphics[width=\textwidth]{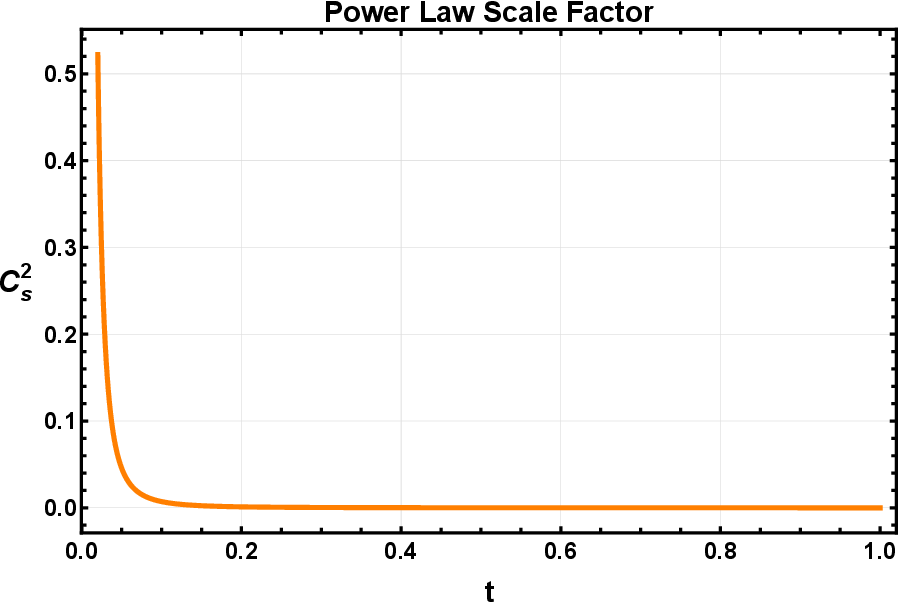}
\end{minipage}
\caption{Variation of squared speed of sound versus cosmic time for power law scale factor model in EPN gravity.}
\end{figure}
\textbf{FIG. 6} illustrate the evolution of \( c_s^2 \) against cosmic time $t$ for the power-law scale factor model in EPN gravity. The figure describes that \( c_s^2 \) remains positive throughout the cosmic evolution, which confirms the absence of Laplacian instabilities and ensures the classical stability of scalar perturbations.
   
\item For the exponential scale factor model given in Eq. (\ref{32}), the field Eqs. (\ref{16}) and (\ref{17}) become as
    \begin{align}\label{5.12}
\rho_{{EPN}} &= \frac{1}{2} c_m \Lambda ^4 y^{2/3} \sqrt[3]{\frac{\Lambda ^4 e^{-2 \lambda  t} \left(e^{\lambda  t}+1\right)^2}{\lambda ^2 M_{pl}^2}}, \\\label{5.13}
p_{{EPN}} &= -\frac{\lambda^2 M_{pl}^2 e^{\lambda  t} \left(3 e^{\lambda  t}+2\right)}{\left(e^{\lambda  t}+1\right)^2}.
\end{align}
From the above Eqs. (\ref{5.12}) and (\ref{5.13}), the squared speed of sound parameter can be given as
\begin{eqnarray}\label{5.14}
C_{s}^{2}=\frac{6 \lambda ^4 M_{pl}^4 e^{3 \lambda  t} \left(2 e^{\lambda  t}+1\right) \left(\frac{\Lambda ^4 e^{-2 \lambda  t} \left(e^{\lambda  t}+1\right)^2}{\lambda ^2 M_{pl}^2}\right)^{2/3}}{c_m \Lambda ^8 y^{2/3} \left(e^{\lambda  t}+1\right)^4}.
\end{eqnarray}
\begin{figure}
\centering
\begin{minipage}[b]{0.5\textwidth}
\includegraphics[width=\textwidth]{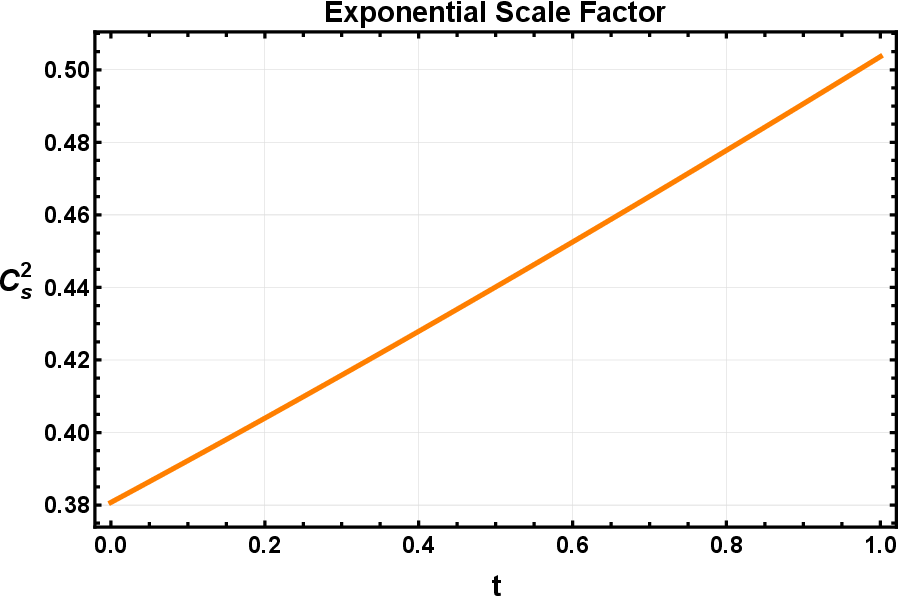}
\end{minipage}
\caption{Variation of squared speed of sound against cosmic time for exponential scale factor model in EPN gravity.}
\end{figure}

\textbf{FIG. 7} represents the variation of the squared speed of sound parameter with cosmic time for the exponential scale factor model in EPN gravity. It is evident that \( c_s^2 \) remains positive and increases monotonically throughout the evolution, ensuring the stability of scalar perturbations. The bounded behavior of \( c_s^2 \) below unity suggests that the propagation of perturbations remains sub-luminal. Therefore, the exponential expansion scenario is dynamically stable and physically viable within the EPN gravity.
    
\item For the modified exponential scale factor model given in Eq. (\ref{37}), the field Eqs. (\ref{16}) and (\ref{17}) becomes as
    \begin{align}\label{5.15}
\rho_{{EPN}} &=\frac{1}{2} c_m \Lambda ^4 y^{2/3} \sqrt[3]{\frac{\Lambda ^4}{M_{pl}^2 (\alpha +2 \beta  t)^2}}, \\\label{5.16}
p_{{EPN}} &= 3 M_{pl}^2 (\alpha +2 \beta  t)^2 \left(-\frac{4 \beta }{3 (\alpha +2 \beta  t)^2}-1\right).
\end{align}
From the above Eqs. (\ref{5.15}) and (\ref{5.16}), the squared speed of sound parameter can be given as
\begin{eqnarray}\label{5.11}
C_{s}^{2}=\frac{18 M_{pl}^4 (\alpha +2 \beta  t)^4 \left(\frac{\Lambda ^4}{M_{pl}^2 (\alpha +2 \beta  t)^2}\right)^{2/3}}{c_m \Lambda ^8 y^{2/3}}.
\end{eqnarray}
\begin{figure}
\centering
\begin{minipage}[b]{0.5\textwidth}
\includegraphics[width=\textwidth]{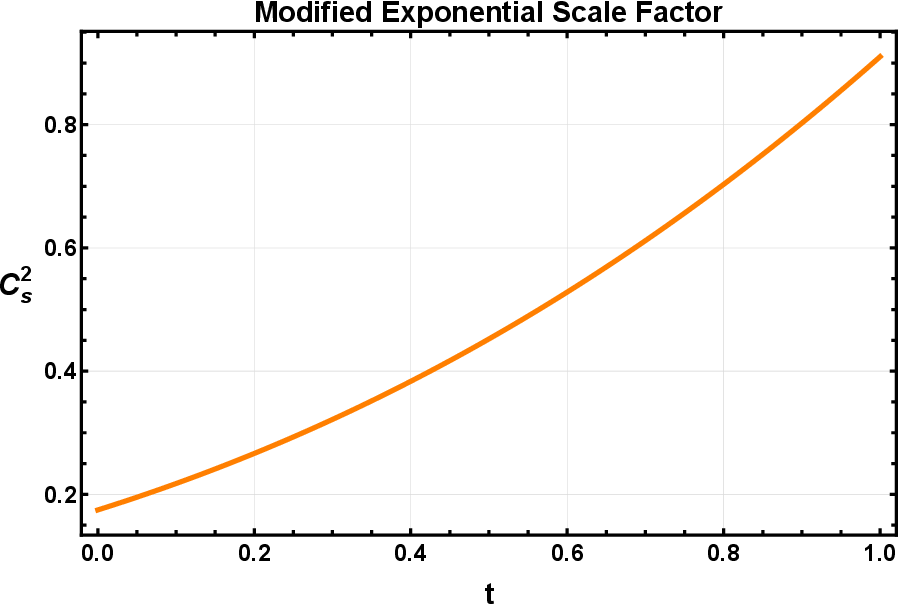}
\end{minipage}
\caption{Variation of squared speed of sound versus cosmic time for modified exponential scale factor model in EPN gravity.}
\end{figure}
\textbf{FIG. 8} gives the evolution of \( c_s^2 \) against cosmic time $t$ for the power-law scale factor model in EPN gravity. The figure describes that \( c_s^2 \) remains positive throughout the cosmic evolution, which confirms the absence of Laplacian instabilities and ensures the stability of scalar perturbations.
\end{itemize}

\begin{table}[h]
\centering
\caption{Comparison of stability criteria used in EPN gravity.}
\begin{tabular}{p{4.5cm} p{5.6cm} p{4.9cm}}
\hline
\hline
~~~~~~~\textbf{Aspect} & ~~~~\textbf{Dynamical Stability} & ~~~\textbf{Sound Speed Stability} \\
\hline
\hline
~~~~~Analysis level & Background cosmological evolution & Linear scalar perturbations \\
~~~~~Method & Phase-space analysis and eigenvalues & $c_s^2=\partial p/\partial\rho$ \\
~~~~~Stability condition & Negative eigenvalues at critical points & $c_s^2 \ge 0$ \\
~~~~~Physical implication & Stable late-time attractor & Absence of Laplacian instability \\
~~~~~Result in EPN gravity & Stable de Sitter phase (trivial case) & $c_s^2>0$ throughout evolution \\
\hline
\hline
\end{tabular}
\end{table}

\section{Results and Discussions}

In this work, we examine the generation of the observed baryon asymmetry through the mechanism of gravitational baryogenesis as well as stability of EPN gravity through dynamical systems and squared speed of sound parameter. The motivation for this work stems from the longstanding problem of explaining the observed matter–antimatter asymmetry in the universe, a phenomenon that standard cosmology and the Standard Model of particle physics alone cannot account for. EPN, as a modified gravity theory, brings new dynamical degrees of freedom, including vector fields, which can play an important role in early universe processes, such as baryon asymmetry generation.

To investigate this, we examined three different cosmological expansion scenarios: power-law, exponential, and modified exponential, each corresponding to distinct phases or behaviors in the early universe's evolution. These models expand our ability to capture a greater spectrum of cosmological dynamics, providing a means to assess the stability of gravitational baryogenesis in a variety of settings. Notably, we incorporated three directional scale factors into the metric to account for anisotropic expansion, while avoiding explicit classification of the geometry to preserve generality. With the modified field equations from the EPN action, we solved the Hubble parameters, Ricci scalar, and other significant quantities for each of the expansion scenarios.

Following this, we used the GB formalism, where the BER arises from the derivative of the Ricci scalar coupled to the baryonic current. This mechanism is very sensitive to how the Ricci scalar evolves, with its behavior being significantly modified in EPN gravity by the vector field and its curvature coupling. The explicit dependence of the Ricci scalar and its time derivative on the scale factors and the Proca field dynamics provided a mechanism for the non-trivial generation of baryon asymmetry.

For all three expansion models, the computed BER is within the observationally accepted range ($\frac{\eta_B}{s} \sim 9.42 \times10^{-11}$), which is in accordance with the CMB and BBN measurements. The models demonstrate flexibility and viability by remaining consistent across a broad range of parameter values. Although we did not directly quantify anisotropic effects using shear scalars or anisotropic stress, the inclusion of three independent scale factors naturally introduces directional dependence in cosmic evolution. This anisotropy, in conjunction with the vector field, refines the gravitational dynamics and offers a more realistic view of the early universe compared to models that assume perfect isotropy. Despite the presence of anisotropic expansion, our results suggest that the baryon asymmetry remains consistent with the observational data. 

Each of the three expansion models leads to successful baryogenesis within observational limits; however, the modified exponential case stands out by offering a more flexible dynamical evolution, which can better accommodate parameter variations while maintaining consistency with empirical data. The exponential scenario, often linked to inflationary dynamics, also produces a stable asymmetry. The power-law scenario acts as a baseline model, showing that even basic anisotropic expansions can facilitate viable baryogenesis within EPN gravity.

In comparison with the previous work, we observe that the phenomenon of GB has been extensively investigated within various modified gravity frameworks. Each theory introduces different geometrical or dynamical degrees of freedom that influence the CPT-violating interaction responsible for generating the baryon asymmetry. Compared with the modified gravity models summarized in Table~\ref{tab:comparison-lit}. Moreover, the key advantage of EPN gravity is that its dynamical vector field naturally produces a non-vanishing time-dependent CPT-violating quantity without requiring specially tuned functions, auxiliary scalar fields, or higher-curvature corrections. In many other frameworks, baryogenesis typically relies on fine-tuned model choices or additional degrees of freedom to keep \( \dot{\mathcal{I}} \neq 0 \) during radiation domination. In contrast, the intrinsic vector geometry couplings in EPN ensure a robust evolution of the baryogenesis-driving invariant across a wide range of cosmological backgrounds, including the scale factors examined in this work. Furthermore, most existing studies of GB focus exclusively on the power law scale factor, whereas our analysis extends the investigation to three distinct cosmological evolutions. This broader treatment demonstrates that EPN gravity can sustain a viable baryogenesis mechanism across different dynamical regimes. This makes EPN a more vibrant, structurally motivated and less fine-tuned environment to achieve the observed baryon asymmetry. Table~\ref{tab:comparison-lit} summarizes a comparative overview of some representative studies in the literature, highlighting the diversity of approaches and the distinctive role played by EPN gravity in our present work.
\begin{table*}[htbp]
\centering
\caption{Comparison among different modified gravity models in the context of gravitational baryogenesis.}
\label{tab:comparison-lit}
\begin{tabular}{p{3.3cm} p{5cm} p{6cm} p{3cm}}
\hline
\hline
\textbf{Gravity Theory} & \textbf{~~~~~~~~Modification} & ~~~\textbf{Main Findings on Baryogenesis} &~~~~~~ \textbf{Reference} \\
\hline
\hline
\( f(G) \) gravity & Inclusion of Gauss--Bonnet invariant \( G \) & Baryogenesis possible via \( \partial_\mu G J^\mu \); links baryon asymmetry to higher-curvature corrections. & ~~~~~~~~~~~~\cite{011} \\
\hline
Non-minimally coupled \( f(R) \) gravity & Curvature-scalar coupling with matter & Effective baryogenesis achieved for specific \( f(R) \) forms. & ~~~~~~~~~~~~\cite{013} \\
\hline
\( f(T,\Theta) \) gravity & Coupling of torsion scalar \( T \) with trace of energy-momentum tensor & Torsion driven baryogenesis consistent with latest observed ratio. & ~~~~~~~~~~~~\cite{015} \\
\hline
\( f(Q,C) \) gravity & Non-metricity-based framework with invariant \( Q \) & Viable baryon-to-entropy ratio consistent with observations. &~~~~~~~~~~~~\cite{016} \\
\hline
\( f(T,\tau) \) gravity & Torsion scalar coupled with trace \( \tau \) of energy-momentum tensor & Observational constraints restrict coupling strength. & ~~~~~~~~~~~~\cite{017} \\
\hline
\( f(R,L_m) \) gravity & Coupling of curvature scalar with matter Lagrangian & Generates asymmetry without extra fields. &~~~~~~~~~~~~\cite{018} \\
\hline
\( f(R,G,T) \) gravity & Combination of curvature, Gauss--Bonnet, and trace terms & Provides enhanced control on baryon-to-entropy ratio. & ~~~~~~~~~~~~\cite{019} \\
\hline
\( f(R,T) \) gravity & Matter-geometry coupling via trace \( T \) & Asymmetry sensitive to matter coupling parameter. & ~~~~~~~~~~~~\cite{023} \\
\hline
Extended teleparallel gravity & Higher-order torsion contributions & Strong baryogenesis for rapidly varying torsion scalar. & ~~~~~~~~~~~~\cite{024} \\
\hline
Ricci inverse gravity & Inverse curvature corrections & Produces viable baryon-to-entropy ratio for late-time expansion. & ~~~~~~~~~~~~\cite{022a} \\
\hline
Einstein-{\AE}ther gravity & Dynamical vector field with preferred frame & Baryogenesis depends on aether coupling; consistent with observations. & ~~~~~~~~~~~~\cite{025} \\
\hline
\( f(T,\phi) \) gravity & Coupled torsion and scalar field model & Reproduces observed baryon asymmetry. & ~~~~~~~~~~~~\cite{025a} \\
\hline
\( f(T,B,T_G,B_G) \) gravity & Extended teleparallel theory with boundary and GB-like terms & Yields consistent baryogenesis with generalized torsional invariants. & ~~~~~~~~~~~~\cite{025b} \\
\hline
\( f(Q,T) \) gravity & Non-metricity and trace coupling & Positive baryon-to-entropy ratio achieved; consistent with Planck constraints. &~~~~~~~~~~~~\cite{026} \\
\hline
\( f(Q) \) gravity & Pure non-metricity formulation & Viable baryogenesis without curvature or torsion; parameter bounds obtained. & ~~~~~~~~~~~~\cite{027} \\
\hline
EPN gravity & Vector-tensor theory with non-minimal coupling between vector field and geometry & Dynamical vector field drives cosmic expansion and CPT violation simultaneously, yielding a unified mechanism for baryogenesis considering three scale factor models. &~~~~~~ This work \\
\hline
\hline
\end{tabular}
\end{table*}

Moreover, the stability analysis of EPN gravity via dynamical system analysis depicts that cosmic evolution naturally progresses from the radiation dominated era towards a stable EPN-dominated de Sitter attractor, confirming a consistent late-time behavior. However, the positivity of squared speed sound parameter for all considered scale factor models indicates the absence of Laplacian instabilities and sub-luminal propagation of scalar perturbations. In general, these results collectively demonstrate that EPN gravity is dynamically and perturbatively stable, strengthening its physical viability as a modified gravity framework.

In summary, we can say that our results establish that EPN gravity provides a consistent and observationally supported approach for explaining the matter–antimatter asymmetry through the GB mechanism. With its ability to yield the correct BER across different expansion models, and without relying on exotic physics, the model presents a promising avenue for future research. Furthermore, the dynamics of the vector field and anisotropic expansion provide novel ways to explore early universe cosmology within the modified gravity context.

\section*{Acknowledgment}

This research is funded by the Science Committee of the Ministry of Science and Higher Education of the Republic of Kazakhstan (Grant No. AP23483654).






\vspace{.25cm}

\begin{thebibliography}{36}

\bibitem{001} P. Arnold and L. McLerran.: Sphalerons, small fluctuations, and baryon-number violation in electroweak theory. $Phys.~ Rev.~ D$ \textbf{36}, 581 (1987).

\bibitem{002} M. Kobayashi and T. Maskawa.: CP-violation in the renormalizable theory of weak interaction. $Prog.~ Theo.~ Phys.$ \textbf{49}, 652 (1973).

\bibitem{003} H. Davoudiasl, R. Kitano, G. D. Kribs, H. Murayama and P. J. Steinhardt.: Gravitational baryogenesis. $Phys.~ Rev.~ Lett.$ \textbf{93}, 201301 (2004).

\bibitem{004} A. Riotto.: Theories of baryogenesis. $ICTP~ summer~ school~ in~ high-energy~ physics~ and~ cosmology$, \textbf{326} (1998).

\bibitem{0044} D. Bödeker, and W. Buchmüller.: Baryogenesis from the weak scale to the grand unification scale. $Rev.~ Mod.~ Phys.$ \textbf{93}, 035004 (2021).

\bibitem{005} A. D. Sakharov.: Violation of CP-invariance, C-asymmetry, and baryon asymmetry of the Universe. In The Intermission Collected Works on Research into the Essentials of Theoretical Physics in Russian Federal Nuclear Center, $Arzamas$, \textbf{16} 84 (1998).

\bibitem{006} A. D. Sakharov.: Baryon asymmetry of the universe. $Soviet~ Phys. Uspekhi$, \textbf{34}, 417 (1991).

\bibitem{007} D. S. Pereira, J. Ferraz, F. S. Lobo and J. P. Mimoso.: Baryogenesis: A Symmetry Breaking in the Primordial Universe Revisited. $Symmetry$ \textbf{16}, 13 (2023).

\bibitem{008} N. Turok.: Physics of the Early Universe: Baryogenesis; Defects; Initial Conditions. Conference Paper: In The primordial universe-L’univers primordial: \textbf{439}, Springer Berlin Heidelberg (2002).

\bibitem{009} M. Trodden.: Electroweak baryogenesis. $Rev.~ Mod.~ Phys.$ \textbf{71}, 1463 (1999).

\bibitem{0010} D. E. Morrissey and M. J. Ramsey-Musolf.: Electroweak baryogenesis. $New~ J.~ Phys.$ \textbf{14}, 125003 (2012).

\bibitem{0011} S. Davidson, E. Nardi and Y. Nir.: Leptogenesis. $Phys.~Rep.$ \textbf{466}, 105 (2008).

\bibitem{0012} W. Buchmüller, R. D. Peccei and T. Yanagida.: Leptogenesis as the origin of matter. $Annu.~ Rev.~ Nucl.~ Part.~ Sci.$ \textbf{55}, 311 (2005).

\bibitem{0013} C. S. Fong, E. Nardi and A. Riotto.: Leptogenesis in the Universe. $Advan.~ High~ Ener.~ Phys.$ \textbf{2012}, 158303 (2012).

\bibitem{0014} W. Buchmüller, P. Di Bari and M. Plümacher.: Leptogenesis for pedestrians. $Annals~ of Phys.$ \textbf{315}, 305 (2005).

\bibitem{0015} J. M. Cline.: Is electroweak baryogenesis dead?. $Philosophical~ Transactions~ of~ the~ Royal~ Society~ A$ \textbf{376}, 20170116 (2018).

\bibitem{0016} M. A. Luty.: Baryogenesis via leptogenesis. $Phys.~ Rev.~ D$ \textbf{45}, 455 (1992).

\bibitem{010} V. K. Oikonomou and E. N. Saridakis.: $f(T)$ gravitational baryogenesis. $Phys.~ Rev.~ D$ \textbf{94}, 124005 (2016).

\bibitem{011} S. D. Odintsov and V. K. Oikonomou.: Gauss–Bonnet gravitational baryogenesis. $Phys.~ Lett. ~B$, \textbf{760}, 259 (2016).

\bibitem{012} M. P. L. P. Ramos and J. Paramos.: Baryogenesis in nonminimally coupled $f (R)$ theories. $Phys.~ Rev.~ D$ \textbf{96}, 104024 (2017).

\bibitem{013} A. Aghamohammadi, H. Hossienkhani and K. Saaidi.: Anisotropy effects on baryogenesis in $f (R)$ theories of gravity. $Mod.~ Phys.~ Lett.~ A$ \textbf{33}, 1850072 (2018).

\bibitem{014} J. Sakstein and A. R. Solomon.: Baryogenesis in Lorentz-violating gravity theories. $Phys.~ Lett.~ B$ \textbf{773}, 186 (2017).

\bibitem{015} S. Rani, N. Azhar and A. Jawad.: $f (T, \Theta)$ gravity and generalized gravitational baryogenesis. $Mod.~ Phys.~ Lett.~ A$ \textbf{37}, 2250056 (2022).

\bibitem{016} M. Usman, A. Jawad and A. M. Sultan.: Compatibility of gravitational baryogenesis in $f(Q, C)$ gravity. $Eur.~ Phys.~ J.~ C$ \textbf{84}, 868 (2024).

\bibitem{017} S. S. Mishra, S. Mandal and P. K. Sahoo.: Constraining $f (T,\tau)$ gravity with gravitational baryogenesis. $Phys.~ Lett.~ B$ \textbf{842}, 137959 (2023).

\bibitem{018} L. V. Jaybhaye, S. Bhattacharjee and P. K. Sahoo.: Baryogenesis in $f (R, L_m)$ gravity. $Phys.~ Dark~ Univ.$ \textbf{40}, 101223 (2023).

\bibitem{019} A. Jawad, M. Usman and M. M. Alam.: Compatibility of Gravitational Baryogenesis with theoretical framework of $f (R, G, T)$ Gravity. $Phys.~ Dark~ Univ.$ \textbf{46}, 101631 (2024).

\bibitem{020} P. C. M. Delgado, M. B. Jesus, N. Pinto-Neto, T. Mourão and G. S. Vicente.: Baryogenesis in cosmological models with symmetric and asymmetric quantum bounces. $Phys.~ Rev.~ D$ \textbf{102}, 063529 (2020).

\bibitem{021} A. Jawad, A. M. Sultan and S. Rani.: Viability of baryon to entropy ratio in modified Hořava–Lifshitz gravity. $Symmetry$ \textbf{15}, 824 (2023).

\bibitem{022} S. Maity and P. Rudra.: Gravitational Baryogenesis in Hořava–Lifshitz gravity. $Mod.~ Phys.~ Lett.~ A$ \textbf{34}, 1950203 (2019).

\bibitem{023} P. K. Sahoo and S. Bhattacharjee.: Gravitational baryogenesis in non-minimal coupled $f (R, T)$ gravity. $Int.~ J.~ Theo.~ Phys.$ \textbf{59}, 1451 (2020).

\bibitem{024} S. Bhattacharjee.: Gravitational baryogenesis in extended teleparallel theories of gravity. $Phys.~ Dark~ Univ.$ \textbf{30}, 100612 (2020).

\bibitem{022a} A. Jawad, A. M. Sultan.: Analysis of baryon to entropy ratio in Ricci inverse gravity. $EPL$ \textbf{138}, 29001 (2022).

\bibitem{025} A. M. Sultan, A. Mushtaq, D. Chou, H. U. Rehman, H. Ashraf, A. U. Awan.: Observational analysis of gravitational baryogenesis constraints in Einstein-Æther gravity. $J.~High~Ener.~Astrophys.$ \textbf{45}, 135 (2025).

\bibitem{025a} A. M. Sultan, A. Mushtaq, A. Jawad, S. Shaymatov, M. S. Saleem.: Gravitational baryogenesis analysis of observationally favored $f(T,\phi)$ model. $Phys.~Dark~Univ.$ \textbf{48}, 101936 (2025).

\bibitem{025b} A. Majeed, S. Rani, N. Azhar, M. M. Alam, S. Shaymatov, A. M. Sultan.: $f(T, B, T_G, B_G)$ gravity tested through gravitational baryogenesis mechanisms. $Phys.~Dark~Univ.$ \textbf{48}, 101957 (2025).

\bibitem{026} S. Bhattacharjee and P. K. Sahoo.: Baryogenesis in $f (Q, T)$ gravity. $Eur.~ Phys.~J.~ C$ \textbf{80}, 289 (2020).

\bibitem{027} S. S. Mishra, A. Bhat and P. K. Sahoo.: Probing baryogenesis in $f(Q)$ gravity. $Europhys.~ Lett.$ \textbf{146}, 29001 (2024).

\bibitem{a} K. Esmakhanova, et al. "Dark energy in some integrable and nonintegrable FRW cosmological models." International Journal of Modern Physics D 20.12 (2011): 2419-2446.

\bibitem{027a} N. Myrzakulov, et al. "Cosmological model of F (T) gravity with fermion fields via Noether symmetry." Journal of Physics: Conference Series. Vol. 1391. No. 1. IOP Publishing, 2019.

\bibitem{027aa} N. Myrzakulov, R. Myrzakulov, and L. Ravera. "Metric-affine Myrzakulov gravity theories." Symmetry 13.10 (2021): 1855.

\bibitem{b} K. Esmakhanova, et al. "Dark energy in some integrable and nonintegrable FRW cosmological models." International Journal of Modern Physics D 20.12 (2011): 2419-2446.

\bibitem{01} A. De Felice, L. Heisenberg, R. Kase, S. Mukohyama, S. Tsujikawa and Y. L. Zhang.: 2016. Cosmology in generalized Proca theories. $JCAP$ \textbf{2016}, 048 (2016).

\bibitem{02} A. De Felice, L. Heisenberg and S. Tsujikawa.: Observational constraints on generalized Proca theories. $Phys.~ Rev.~ D$ \textbf{95}, 123540 (2017). 

\bibitem{03} S. Nakamura, R. Kase and S. Tsujikawa.: Cosmology in beyond-generalized Proca theories. $Phys.~ Rev.~ D$ \textbf{95}, 104001 (2017).

\bibitem{04} C. de Rham  and V. Pozsgay.: New class of Proca interactions. $Phys.~ Rev.~ D$, \textbf{102}, 083508 (2020).

\bibitem{05} C. de Rham, S. Garcia-Saenz, L. Heisenberg and V. Pozsgay.: Cosmology of Extended Proca-Nuevo. $JCAP$ \textbf{03}, 053 (2022).

\bibitem{06} C. de Rham, L. Heisenberg, A. Kumar and J. Zosso.: Quantum stability of a new Proca theory. $Phys.~ Rev.~ D$, \textbf{105}, 024033 (2022).

\bibitem{07} F. K. Anagnostopoulos and E. N. Saridakis.: Observational constraints on extended Proca-Nuevo gravity and cosmology. $JCAP$ \textbf{2024}, 051 (2024).

\bibitem{08} L. Sudharani, N. S. Kavya and V. Venkatesha.: Unveiling the effects of coupling extended Proca-Nuevo gravity on cosmic expansion with recent observations. $Mon.~Not.~Roy.~Astro.~Soc.$ \textbf{535}, 1998 (2024).

\bibitem{09} N. S. Kavya, L. Sudharani and V. Venkatesha.: Constraining extended Proca-Nuevo theory through big bang nucleosynthesis. $Gen.~ Rel.~ Grav.$ \textbf{57}, 1 (2025).

\bibitem{045} C. de Rham, G. Gabadadze and A. J. Tolley.: Resummation of massive gravity. $Phys.~ Rev.~ Lett.$ \textbf{106}, 231101 (2011).

\bibitem{046} C. de Rham.: Massive gravity. $Living~ Rev.~ Rel.$ \textbf{17}, 7 (2014).













\bibitem{1} C. L. Bennett, et al.: WMAP Collaboration, The Microwave Anisotropy Probe Mission. $Astrophys.~J.~ Suppl.$ \textbf{148}, 1 (2003).

\bibitem{2} S. Burles, K. M. Nollett, M. S. Turner.: What is the big-bang-nucleosynthesis prediction for the baryon density and how reliable is it? $Phys.~ Rev.~ D$ 63, 063512 (2001).

\bibitem{45} S. H. S. Alexander, M. E. Peskin and M. M. Sheikh-Jabbari.: Leptogenesis from gravity waves in models of inflation. $Phys.~ Rev.~ Lett.$ \textbf{96}, 081301 (2006).

\bibitem{46} S. Mohanty, A. R. Prasanna and G. Lambiase.: Leptogenesis from spin-gravity coupling following inflation. $Phys.~ Rev.~ Lett.$ \textbf{96}, 071302 (2006).

\bibitem{41} P. A. R. Ade, et al.: Planck Collaboration: Planck intermediate results-XLV. Radio spectra of northern extragalactic radio sources. $Astron.~ Astrophys.$ \textbf{594}, 103 (2016).

\bibitem{42} E. W. Kolb, M. S. Turner., The Early Universe. $Front.~ Phys.$ \textbf{69}, 1 (1990).

\bibitem{43} H. Davoudiasl, et al.: Gravitational baryogenesis. $Phys.~ Rev.~ Lett.$ \textbf{93}, 201301 (2004).

\bibitem{44} K. Nozari and F. Rajabi.: Baryogenesis in $f (R, T)$ Gravity. $Comm.~ Theo.~ Phys.$ \textbf{70}, 451 (2018).

\bibitem{055} A. Dolgov, V. Halenka and I. Tkachev.: Power-law cosmology, SN Ia, and BAO. $JCAP$ \textbf{10}, 047 (2014).

\bibitem{056} S. Nojiri, S. D. Odintsov and V. K. Oikonomou.: The bounce universe history from unimodular $F(R)$ gravity. $Phys.~ Rev.~ D$ \textbf{93}, 084050  (2016).

\bibitem{v} K. Y. Kim, H.W. Lee and Y.S. Myung, Phys. Lett. B 660 (2008) 118.

\end{thebibliography}
\end{document}